\documentclass{PoS}

\usepackage{cleveref}
\renewcommand{\cref}{\Cref}
\usepackage{epstopdf}
\newcommand{\veritas}{\textsc{Veritas}}
\newcommand{\magic}{\textsc{Magic}}
\newcommand{\GeV}{\mathrm{GeV}}

\usepackage{url}
\usepackage[percent]{overpic}
\usepackage{transparent}

\title{\veritas\ Observations under Bright Moonlight}

\ShortTitle{\veritas\ Observations under Bright Moonlight}

\author{\speaker{Sean Griffin} for the VERITAS collaboration\thanks{veritas.sao.arizona.edu}\\
        McGill University\\
        E-mail: \email{griffins@physics.mcgill.ca}}

\abstract{The presence of moonlight is usually a limiting factor for imaging atmospheric Cherenkov telescopes due to the high sensitivity of the camera photomultiplier tubes (PMTs). In their standard configuration, the extra noise limits the sensitivity of the experiment to gamma-ray signals and the higher PMT currents also accelerates PMT aging. Since fall 2012, observations have been carried out with \veritas\ under bright moonlight (Moon illumination > 35\%), in two observing modes, by reducing the voltage applied to the PMTs and with UV bandpass filters, which allow observations up to $\sim80\%$ Moon illumination resulting in 29\% more observing time over the course of the year. In this presentation, we provide details of these new observing modes and their performance relative to the standard \veritas\ observations.}

\FullConference{The 34th International Cosmic Ray Conference,\\
		30 July- 6 August, 2015\\
		The Hague, The Netherlands}

\begin{document}

\section{A Brief History of Observing Under Moonlight}

For imaging atmospheric Cherenkov telescopes (IACTs), the increased sky brightness due to the Moon has always been a limiting factor in the amount of observing time available. Historically, IACTs were limited to observing during moonless nights.  To circumvent these limitations and increase the duty cycle of the telescopes, a number of collaborations have experimented with different methods of observing under moonlight.

The \textsc{Artemis} experiment, designed to measure the shadowing of cosmic rays by the Moon at TeV energies \cite{ARTEMIS90,ARTEMIS01} required pointing the telescope 1-2$^\circ$ from the Moon and was heavily influenced by triggers due to moonlight photons. Both \textsc{Artemis} and \textsc{Whipple} experimented with the use of UV filters to increase the telescope's sensitivity during moonlight. For \textsc{Whipple}, the filters were able to increase the duty cycle of the experiment, albeit with a substantially increased energy threshold \cite{whippleMoon,whippleFilter}. In the case of \textsc{Artemis}, this improved the sensitivity of the experiment to air showers by a factor of three \cite{ARTEMIS00}. The \textsc{Whipple} collaboration also experimented with the use of solar-blind cameras \cite{whippleSolarBlind}.

The \textsc{Hegra} experiment pioneered regular observations under moderate moonlight using reduced photomultiplier tube (PMT) gains \cite{hegra}. The \magic\ collaboration regularly observes during both twilight and moderate moonlight using the same method \cite{magic}.

In this work, we discuss two new observing modes for \veritas\ based on both the reduced camera gain and filter methods of coping with moonlight. 

\section{The \veritas\ Observing Strategy}

\veritas\ employs a ``safety threshold'' to prevent the PMTs from being damaged by large amounts of light. At all times, the mean PMT currents across each of the cameras must be less than $15~\mathrm{\mu A}$. When this threshold is exceeded, the observers are instructed to power down the cameras. Prior to the advent of the Bright Moonlight Program, this limited \veritas\ observations in the standard/nominal configuration (NOM) to when the Moon was less than $35\%$ illuminated. This is a guideline more than a rule; PMT currents are a function of Moon illumination, the position of the Moon in the sky, and the telescope pointing position. Furthermore, if there are clouds in the sky, scattered moonlight can brighten the sky to the point where additional wear on the PMTs become an issue. 

\begin{figure}[hbtp]
\centering
\begin{overpic}[width=0.7\textwidth,tics=10]{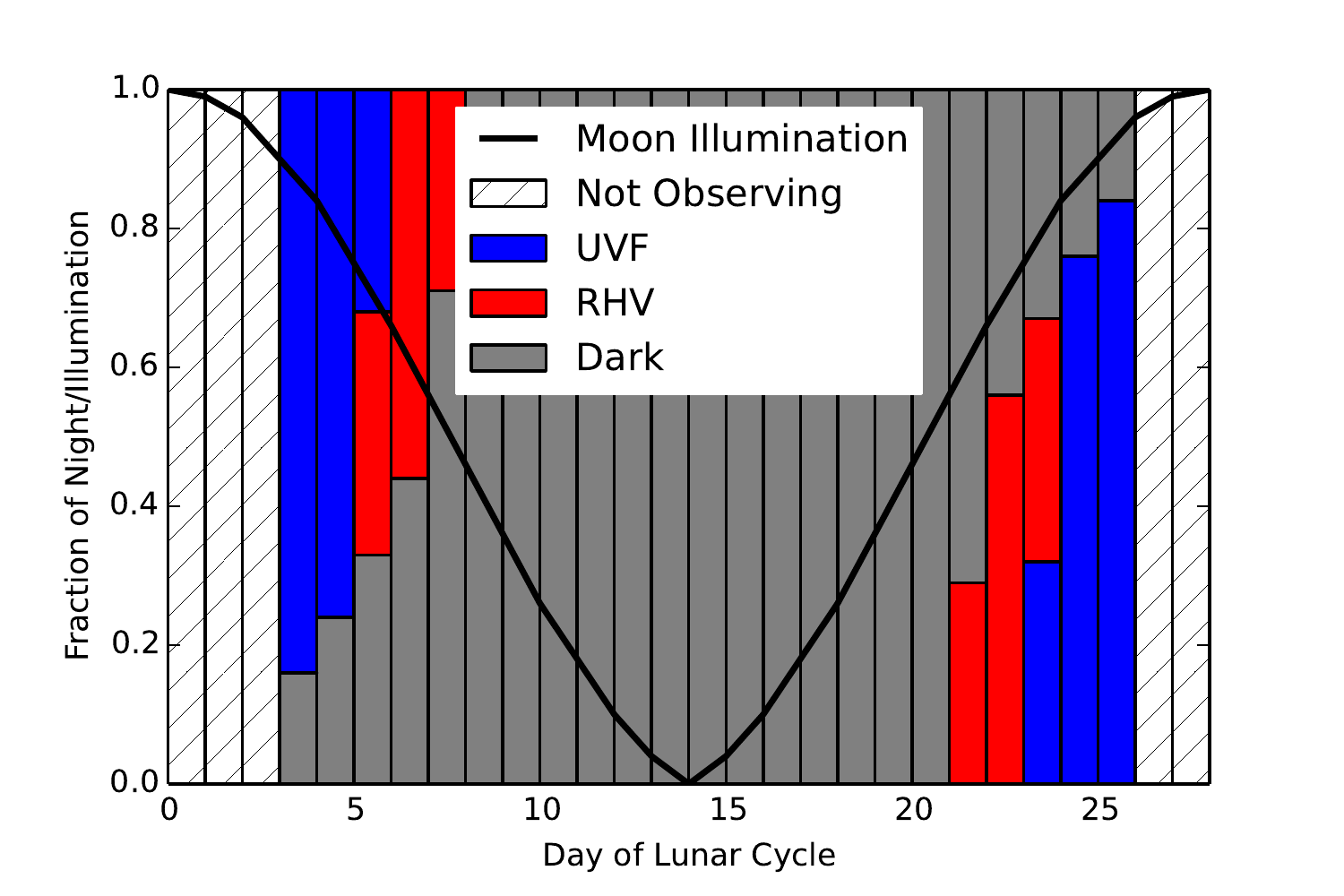}
 \put (33.5,30) {\transparent{1.0}\textcolor{white}{VERITAS - ICRC 2015}}
\end{overpic}
\caption{Breakdown of a typical observing month for \veritas. Note that this strategy is a guideline more than a rule; observing modes are chosen based on the PMT currents at the time of observing, rather than at fixed intervals. The observing pattern at the end of the lunar cycle is inverted with respect to the start; this is due to the fact that near the start, the Moon rises partway through the observing night. Conversely, near the end of the dark run, at the start of observing the Moon is above the horizon and subsequently sets as the night progresses.}
\label{fig:ObservingMonth}
\end{figure}

In order to bypass this restriction, two new observing modes have been developed. The first is the reduced high voltage (RHV) mode which allows \veritas\ to operate when the Moon is $\sim~35-65\%$ illuminated. The second is the use of UV-bandpass filters (UVF) to block out the majority of the additional night sky background (NSB) photons present due to the Moon. In principle, the use of the filters allows \veritas\ to operate through the full Moon, but the period closest the full Moon is typically used for telescope maintenance instead of science observations. The breakdown of a typical observing month is given in \cref{fig:ObservingMonth}.

\subsection{Observing with Reduced High Voltage}

In the RHV observation mode, the PMT voltages are reduced to 81\% of their standard values. The \veritas\ PMTs are nominally operated at an absolute gain of $2\times 10^5$, and the RHV settings reduce this by a factor of $\sim 3.2$. The effect is fewer electrons bombarding the last dynode of the PMT, which reduces the cumulative damage to the last dynode and thus the aging of the PMT.

\subsection{Observing with UV Filters}

Once the Moon is more than $65\%$ illuminated, the sky is too bright to observe using the RHV strategy without damaging the PMTs.  In the Johnson $U$- and $B$- bands \cite{johnson}, the difference in the night sky brightness between the new and full Moon is a factor of $\sim 100$ and $\sim 19$, respectively (see \cite{moonBrightnessRatio} and references therein). While it is possible to decrease the PMT gains further, which would protect the PMT dynodes from radiation damage, it does nothing to protect the photocathode from the constant bombardment of NSB photons. 

\begin{figure}[hbtp]
\centering
\begin{overpic}[angle=270,width=0.5\textwidth,tics=10]{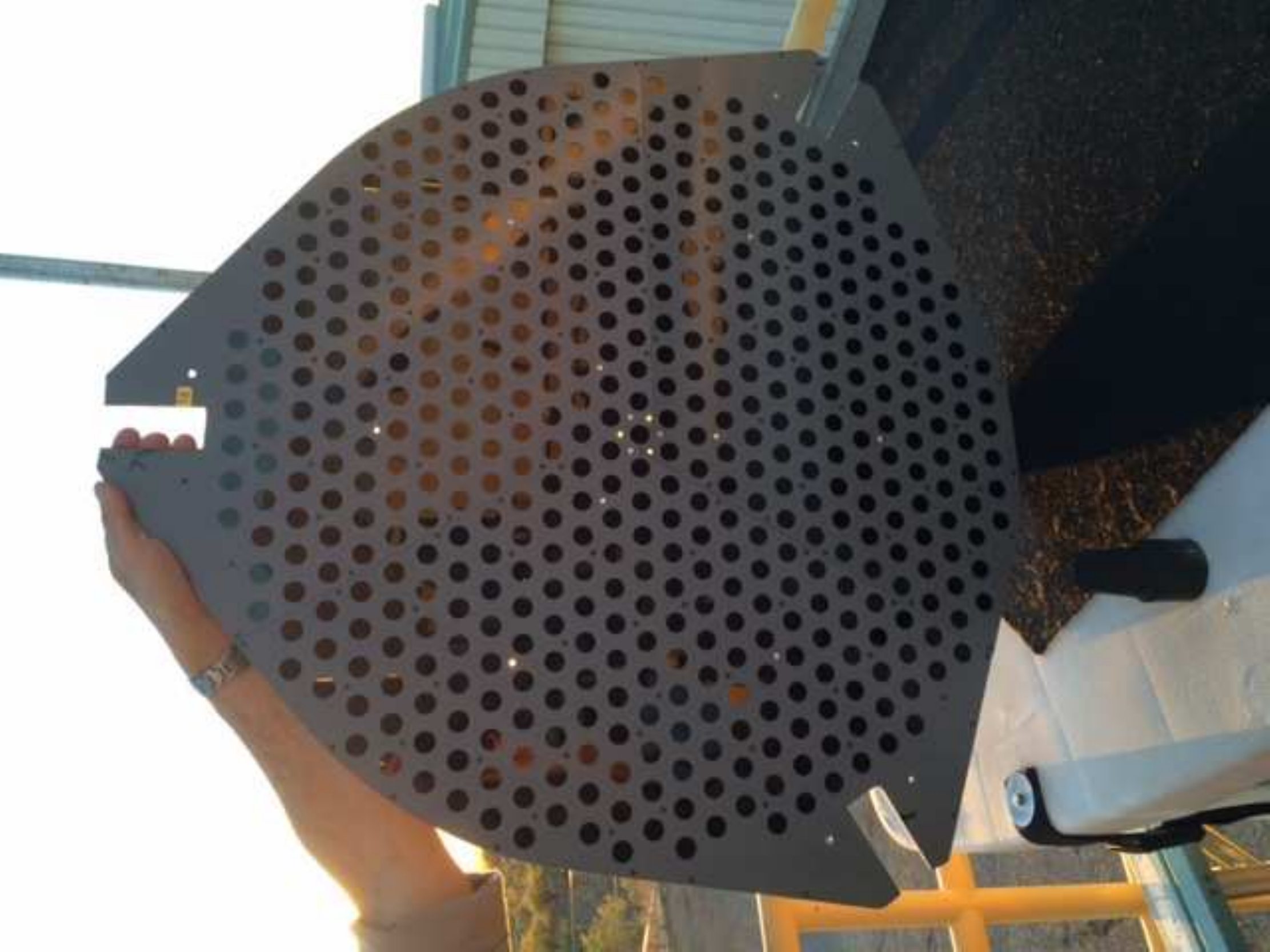}
 \put (37.5,5) {\transparent{1.0}\textcolor{white}{VERITAS - ICRC 2015}}
\end{overpic}
\caption{A UV filter plate. Each black circle is an individual filter. Each filter plate is installed between the Winston light-cone assembly and the PMTs.}
\label{fig:UVFplates}
\end{figure}

\begin{figure}[hbtp]
\centering
\begin{overpic}[width=0.7\textwidth,tics=10]{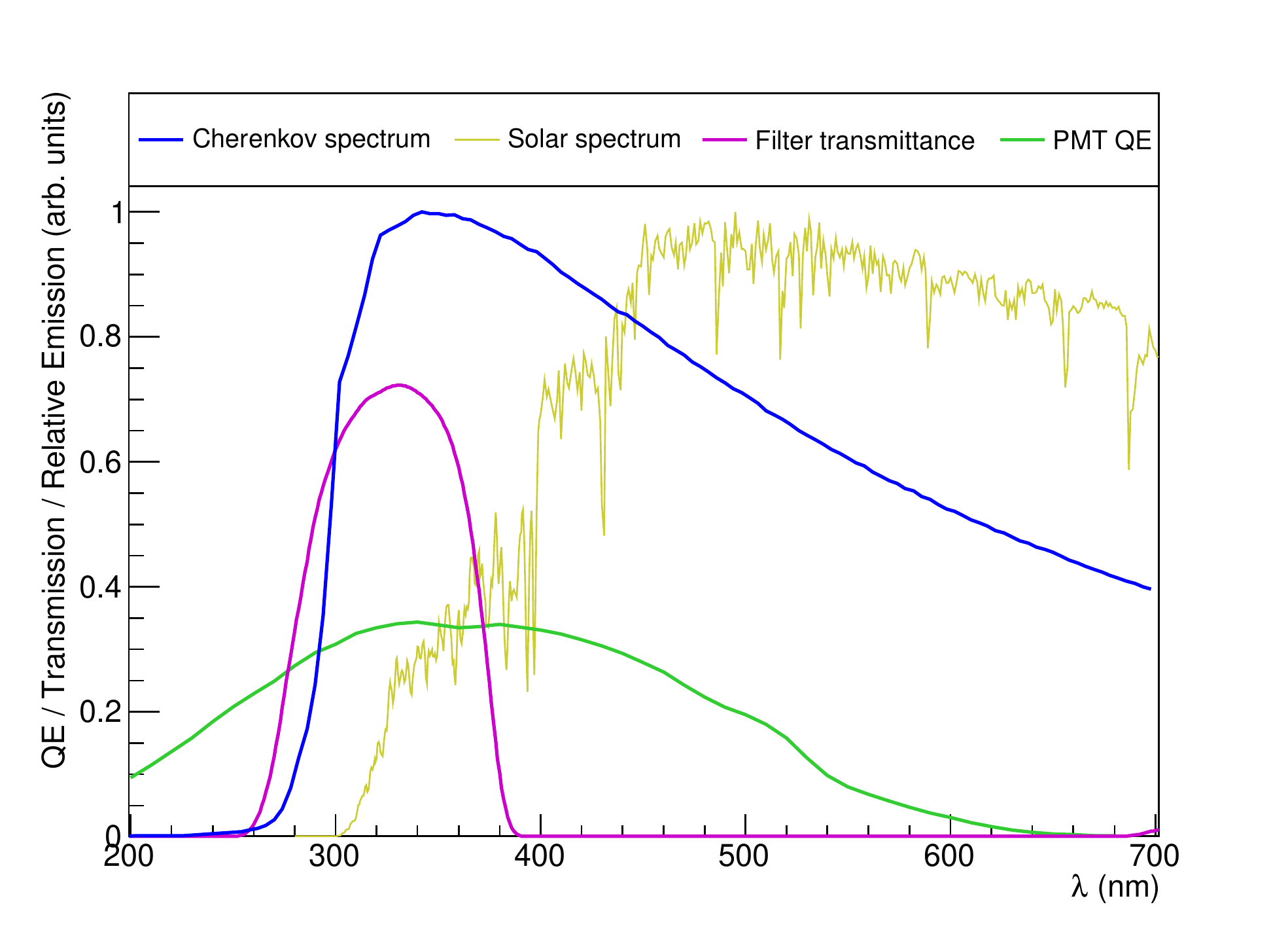}
 \put (50,22.5) {\transparent{1.0}\textcolor{red}{VERITAS - ICRC 2015}}
\end{overpic}
\caption{Transmission spectrum for the \veritas\ filters alongside the solar spectrum from \cite{solarSpec}, Cherenkov spectrum for a $500~\GeV$ gamma ray and the \veritas\ PMT quantum efficiency. $96\%$ of the Moonlight (solar) photons are rejected, but the filters still transmit 17\% of the Cherenkov spectrum over the range shown here.}
\label{fig:UVFSpec}
\end{figure}

In order to overcome this limitation, UV bandpass filters are used to reduce the number of photons hitting the PMT face. The \veritas\ ``filter plates'', shown in \cref{fig:UVFplates}, are made of 499 individual 3-mm-thick SCHOTT UG-11 filters \cite{SCHOTTUG11}.  The transmission spectrum of the filters is shown in \cref{fig:UVFSpec}. To first order, the lunar spectrum is simply reflected sunlight which is then scattered in the Earth's atmosphere \cite{moonModel,moonSpectrum}. Thus, a solar spectrum is also shown in the same figure alongside the Cherenkov spectrum for a $500~\GeV$ gamma-ray at ground level. 

The camera of each \veritas\ telescope is equipped with 499 Winston-type light concentrators \cite{winston} which reduce the dead space between PMTs (thus increasing their effective collection area) and reject photons arriving at the focal plane that do not come from the reflector. When installed, the filter plates are ``sandwiched'' between the PMTs and the Winston cones. The individual filters are larger than the exit apertures of the Winston cones, so the geometry of the setup does not prevent photons from reaching the PMTs.

\veritas\ uses small filters instead of one large filter for a number of reasons. Firstly, it is simpler to manufacture small filters and more cost-effective to purchase them in bulk. Secondly, in the case of an accident while installing the plates, any damaged filters can be replaced for a nominal fee, whereas replacing one large piece of glass would be prohibitively expensive. 

It takes  approximately $60~\mathrm{minutes}$ for a three-person observing crew to install (or uninstall, depending on the night) the UV filters on the array.  In order to maximise the amount of time spent observing in the higher sensitivity RHV mode the changeover is conducted during UVF time.  For nights which start with UVF mode the filters are installed during the day and removed when the Moon drops below an elevation whereby, after the filter removal process, observations can be conducted in RHV mode.  Conversely, if observations end in UVF mode the filters are installed  after the currents exceed the limit for RHV and the filters are then removed during the day.

\section{Results}

In order to verify the detector model used in RHV and UVF data analysis, data from observations of the Crab nebula made while in these modes were analysed and compared to a subset of the \veritas\ nominal configuration data set. These results were produced using the standard \veritas\ analysis pipeline \cite{veritasAnalysis}. The NOM data set was selected to be data taken within a few months of when the RHV and UVF data sets were taken in order to help minimize systematic differences between the data sets.

For nominal configuration data, the sensitivity of \veritas\ is $35-40 \sigma/\sqrt{\mathrm{hr}}$ for a Crab-like source at $70^\circ$ elevation; further details on the sensitivity of \veritas\ in this mode can be found in \cite{nahee}. For the analysis presented here, data were analysed using standard \veritas\ gamma-hadron separation cuts, optimized \emph{a priori} on the Crab nebula.  The energy threshold for NOM data set presented in this work is $\sim 140~\GeV$. Note that \veritas\ cuts are optimized for sensitivity and not energy threshold. 

The analysis of RHV Crab data was performed using similar cuts but with a smaller threshold on the image brightness to account for the reduced pulse heights. The resulting sensitivity is $37\sigma/\sqrt{\mathrm{hr}}$,  with a slightly higher mean energy threshold of $\sim 160~\GeV$. 

\begin{figure}[hbtp]
\centering
\begin{overpic}[width=0.7\textwidth,tics=10]{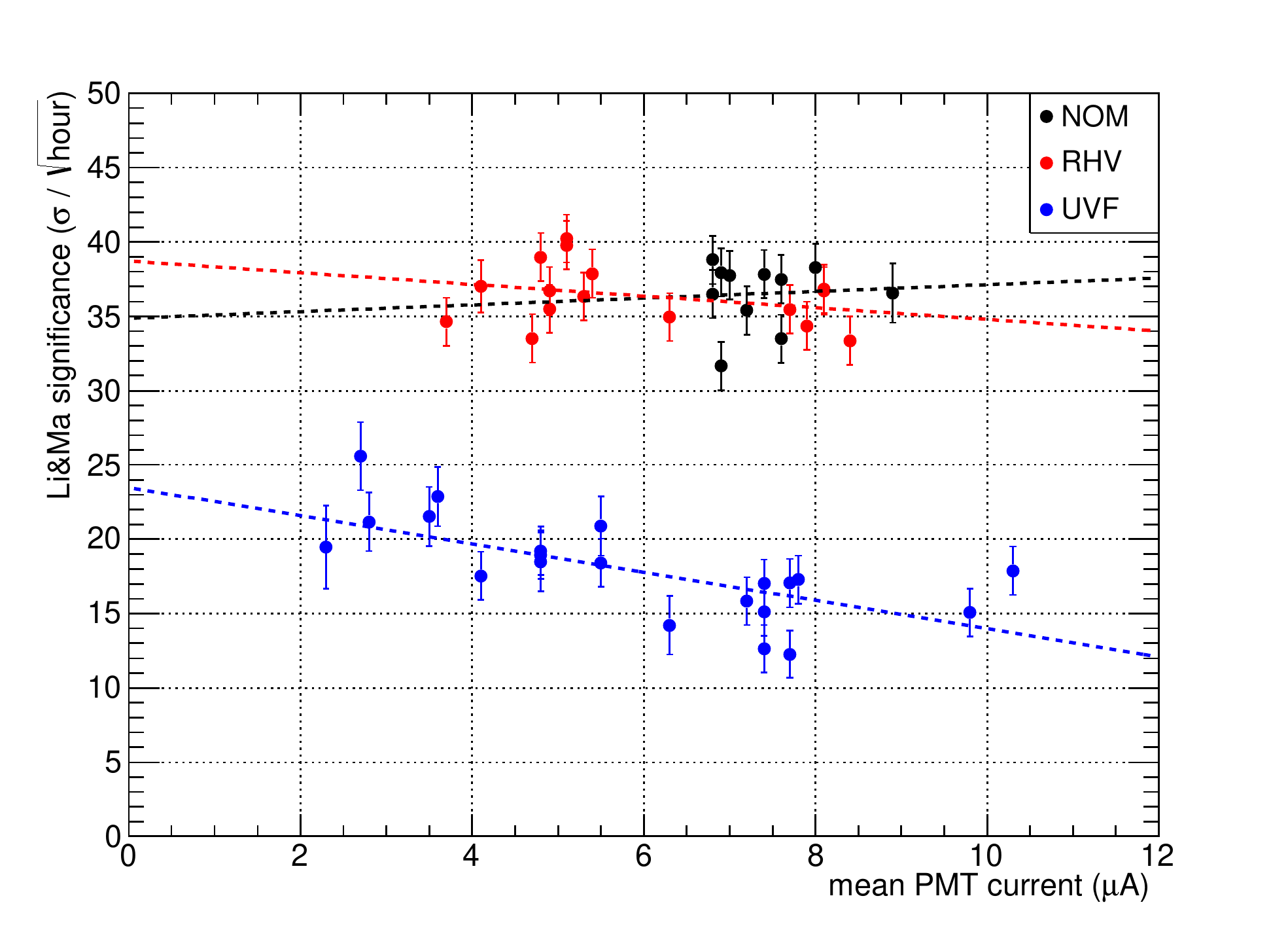}
 \put (52.5,10) {\transparent{1.0}\textcolor{red}{VERITAS - ICRC 2015}}
\end{overpic}
\caption{Run-wise sensitivity of the three different data sets presented in this work. All the runs were taken at elevations above $65^\circ$. The three lines are fits to indicate the general trend; the fit to the NOM data set has a slope which is consistent with zero.}
\label{fig:SensitivityCorrelation}
\end{figure}

A good measure of the robustness of the observing mode against increased noise is the correlation between sky brightness (which is directly proportional to the NSB rate), and the run-by-run sensitivity. A good proxy for the sky brightness is the mean PMT current; this is because for typical light levels the signal produced by a PMT is linearly proportional to the number of photons arriving at the PMT's photocathode. The run-wise sensitivity versus mean camera current has been plotted in \cref{fig:SensitivityCorrelation}. There is a small anti-correlation for the RHV observing mode (correlation coefficient $-0.3$) indicating that the sensitivity is dependent on the sky brightness, but not heavily. For the NOM analysis, intuitively one expects the sensitivity to decrease with increasing sky brightness, but the spread in the PMT currents for the NOM data shown here is too small to quantify the anti-correlation.

The UVF Crab data have a significantly lower sensitivity, $18 \sigma/\sqrt{\mathrm{hr}}$ on average, and has a strong anti-correlation between sensitivity and the sky brightness (correlation coefficient of $-0.7$). The analysis energy threshold for the UVF observing mode is also higher, ranging from $250~\GeV$ to $400~\GeV$ depending on the sky brightness for similar zenith angles. This means that care must be taken when selecting targets for observations with UV filters; this observing mode is best suited for strong, hard-spectrum sources (e.g. pulsar wind nebulae or hard-spectrum blazars). 

\begin{figure}[hbtp]
\centering
\begin{overpic}[width=0.8\textwidth,tics=10]{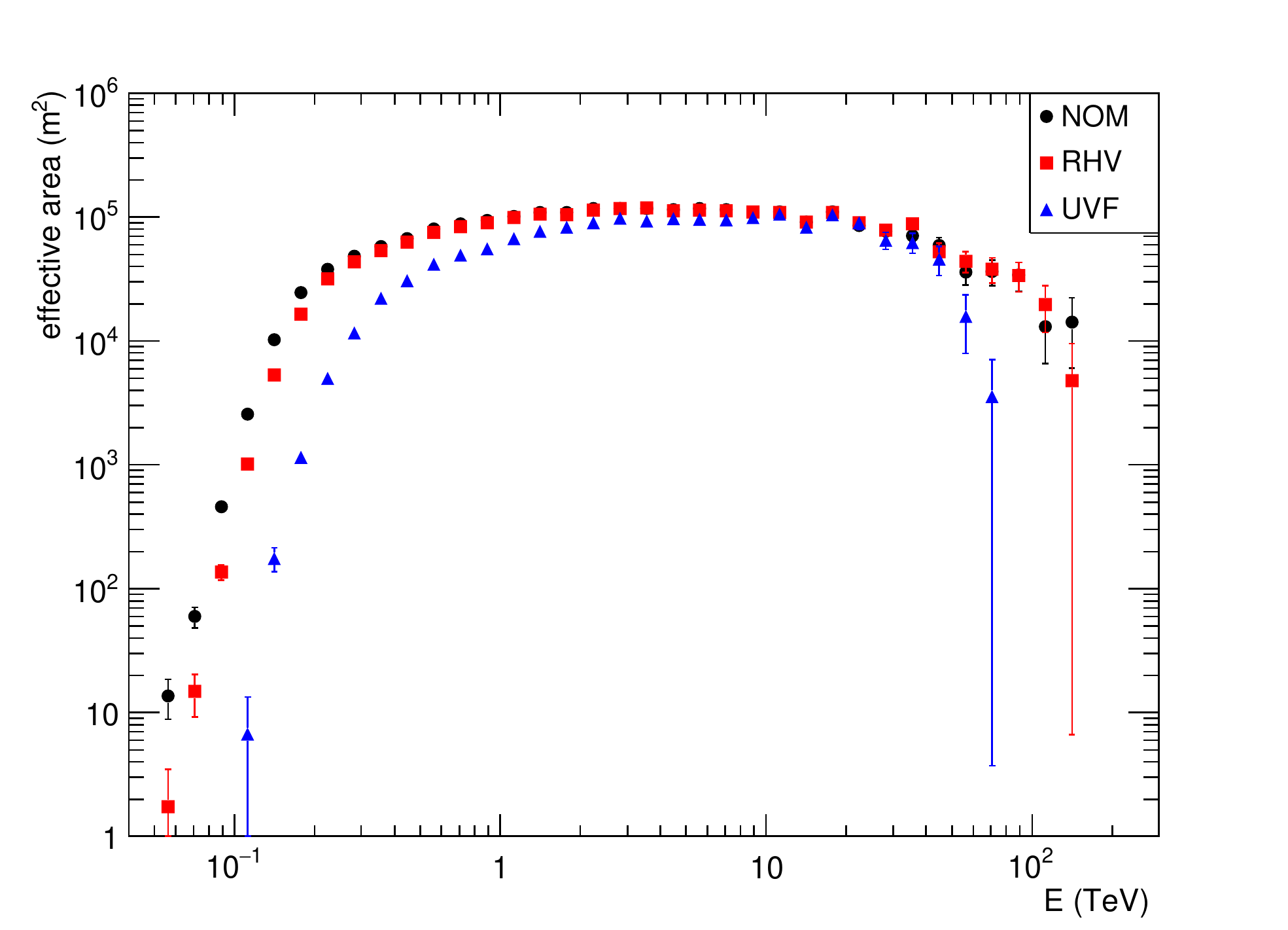}
 \put (52.5,10) {\transparent{130}\textcolor{red}{VERITAS - ICRC 2015}}
\end{overpic}
\caption{Example effective areas for the three observing modes at $20^\circ$ zenith angle. For UVF, the cutoff at high energies is likely due to limited statistics; only half as many showers were used in generating the UVF effective areas as were used for NOM and RHV. This does not affect the analysis here since the highest UVF spectral point is well below this limit.}
\label{fig:3EA}
\end{figure}

\begin{figure}[hbtp]
\centering
\begin{overpic}[width=0.8\textwidth,tics=10]{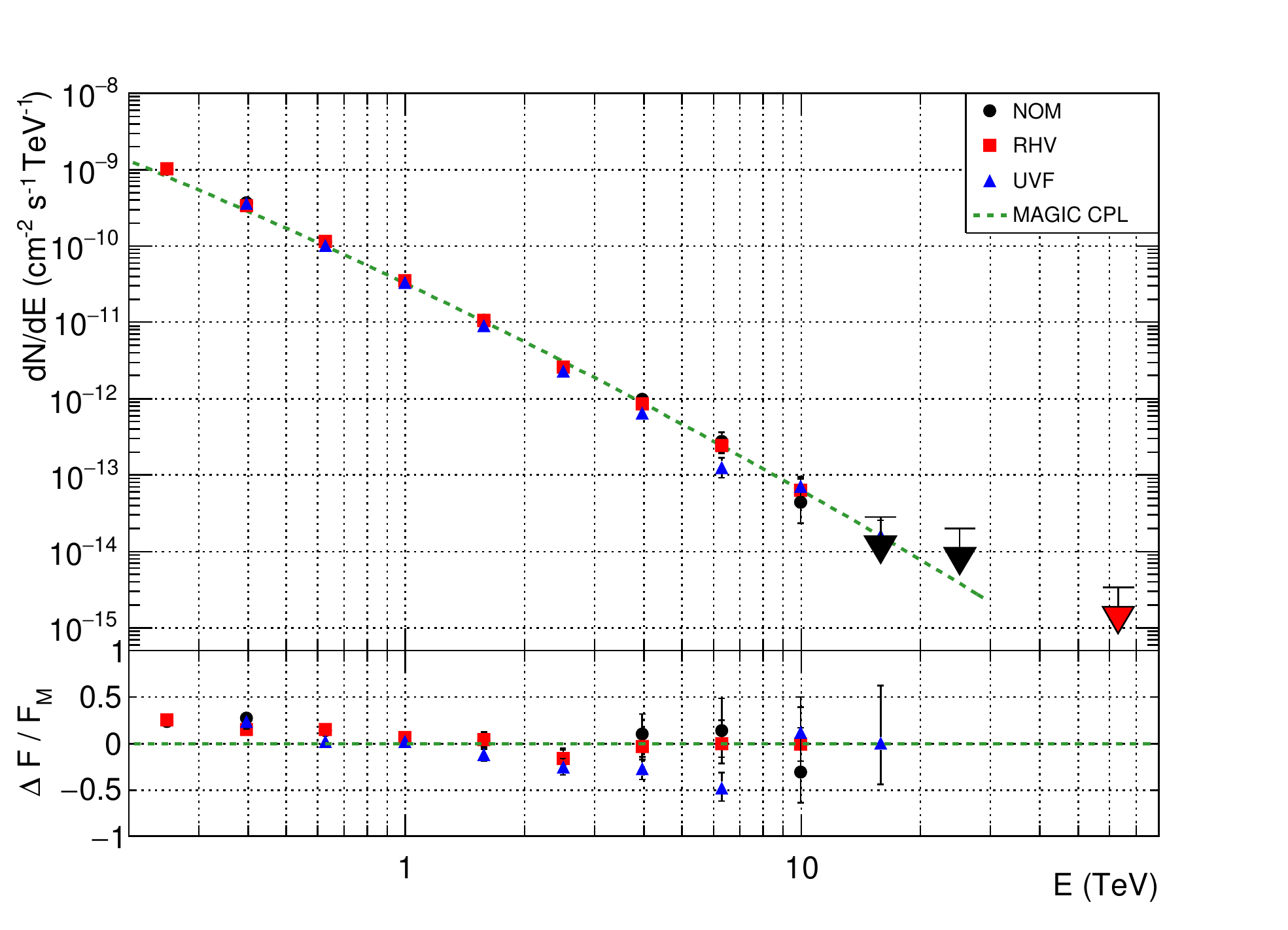}
 \put (52.5,42.5) {\transparent{1.0}\textcolor{red}{VERITAS - ICRC 2015}}
\end{overpic}
\caption{\textbf{Top}: Crab nebula spectra resulting from the NOM, RHV, and UVF data sets. The curved power law from \cite{magicSpec} is also indicated. The upper limits are at the 95\% confidence level. \textbf{Bottom}: Residuals of the three data sets from the curved power law from \cite{magicSpec}. The upper limits have intentionally been omitted for clarity.}
\label{fig:CrabSpectra}
\end{figure}

An example of effective areas used to analyse data from the new observing modes is shown in \cref{fig:3EA} and the corresponding energy spectra from data taken on the Crab nebula are shown in \cref{fig:CrabSpectra}. The curved power law from the \magic\ collaboration \cite{magicSpec} is shown for reference. The three data sets agree very well below 4~TeV; the UVF data set has one spectral point that is statistically inconsistent with the others. This is likely due to poor statistics at higher energies, as it is only this spectral bin that is in disagreement. The data presented here are the result of $5.3$ hours of observing in the standard configuration, $7.8$ hours of RHV data, and $8.9$ hours of UVF data.

The bright moonlight observing mode systematic uncertainties due to mirror reflectivity, uncertainty in the atmosphere, and telescope optical point spread function are the same as those reported for standard observations. The current overall systematic uncertainty for the bright moonlight modes is under investigation, but it is conservatively estimated to be 30\% on the flux normalization and 0.3 on the spectral index. This is 50\% more than the standard \veritas\ systematic uncertainties.
\section{Outlook}

The first science product of the \veritas\ Bright Moonlight Program was the detection of the blazar \emph{1ES~1727+502} in a flaring state during RHV observations \cite{1727}. This was the first detection of variability in this object. This is a flare that otherwise would have gone unnoticed, thus demonstrating one of the benefits of increased observing time. Another interesting program, made possible entirely of by the Bright Moonlight Program is an attempt to measure the positron fraction by observing the cosmic ray shadow due to the Moon \cite{moonShadow}. 

Two new observing modes for \veritas\ have been presented. The RHV mode provides a $13\%$ boost in yearly exposure above $250~\GeV$ and the UVF mode provides a $16\%$ boost in yearly exposure above $500~\GeV$. This additional observing time has been used to increase the livetime of the experiment, allowing for deeper exposures, triggering or following up on astrophysical transient events, and the pursuit of new science goals. 

\subsubsection*{Acknowledgments}
\footnotesize
This research is supported by grants from the U.S. Department of Energy Office of Science, the U.S. National Science Foundation and the Smithsonian Institution, and by NSERC in Canada. We acknowledge the excellent work of the technical support staff at the Fred Lawrence Whipple Observatory and at the collaborating institutions in the construction and operation of the instrument. The \veritas\ Collaboration is grateful to Trevor Weekes for his seminal contributions and leadership in the field of VHE gamma-ray astrophysics, which made this study possible.

\end{document}